# An Innovative Wireless Cardiac Rhythm Management (*iCRM*) System

Gabriel E. Arrobo†, Calvin A. Perumalla†, Stanley B. Hanke†, Thomas P. Ketterl†, Peter J. Fabri‡, and Richard D. Gitlin†

† Department of Electrical Engineering, ‡ Department of Industrial Engineering and College of Medicine
University of South Florida
Tampa, Florida, U.S.A.
{garrobo, calvin4, stanleyh}@mail.usf.edu, ketterl@usf.edu, pfabri@health.usf.edu, richgitlin@usf.edu

*Abstract*—In this paper, we propose a wireless *Communicator* to manage and enhance a Cardiac Rhythm Management System. The system includes: (1) an on-body wireless Electrocardiogram (ECG), (2) an Intracardiac Electrogram (EGM) embedded inside an Implantable Cardioverter/Defibrillator, and (3) a Communicator (with a resident Learning System). The first two devices are existing technology available in the market and are emulated using data from the Physionet database, while the *Communicator* was designed and implemented by our research team.

The value of the information obtained by combining the information supplied by (1) and (2), presented to the Communicator, improves decision making regarding use of the actuator or other actions. Preliminary results show a high level of confidence in the decisions made by the Communicator. For example, excellent accuracy is achieved in predicting atrial arrhythmia in 8 patients using only external ECG when we used a neural network.

## I. INTRODUCTION

The electrocardiogram (ECG/EKG) is an electrical representation of the polarization cycle of the heart as read by electrodes placed at strategic points on the skin surface. The heart is stimulated by the natural pacemaker of the heart, which applies an electrical signal to the heart such that the muscles of the heart contract. The heart's electrical signals set the rhythm of the heartbeat. An ECG records the heart's electrical activity and indicates (1) heart rate, (2) whether the rhythm of the heartbeat is steady or irregular.

Conventionally, 10 electrodes are used to display the ECG and the combined output is known as the 12-lead ECG because 12 distinct signals are recorded and displayed. These signals provide a 3 dimensional (spatial) view of the heart's electrical activity to the physician. With the advancement in technology, a smaller version of the 12-lead ECG that uses 2 or 3 leads has been realized, that can be attached to the human body, and can continuously monitor the heart's electrical activity. This type of ECGs is called an ambulatory ECG, or on-body wireless ECG, and provides only a one or two- dimensional view of the heart's electrical activity. It is hypothesized that by combining the information supplied by ECGs and EGMs, both of which measure the electrical activity of the heart, the value of this combined information presented to the *Communicator* and/or the physician will increase and improve decision-making, that in turn benefits the accuracy of the actuator and other actions. The presentation and analysis of this data, in real time, is facilitated by the *Communicator*. Note that the information that can be provided to the physician by the 12-lead ECG is much more comprehensive than that of the wireless ECG.

In a similar fashion, an Intracardiac EGM, embedded in a pacemaker device, an Implantable Cardioverter Defibrillator (ICD) or Cardiac Resynchronization Therapy (CRT) device monitors the heart for possible arrhythmias [1]. The implanted device is surgically placed in the chest close to the collarbone. Catheters are extended into the heart from the device and come into contact with cardiac tissue. The EGM sensor views the heart's polarization cycle in one dimension. The implanted device processes and analyzes this information to monitor the heart's electrical activity. Based on this information, the device, through an actuator, applies an electrical signal to the heart to bring it out of arrhythmia and back to normal heart rhythm.

The main difference between the ECG and the EGM is that the surface ECG provides a high-level view of the heart and the EGM is a fine-grain view of the heart. The surface ECG in essence has a "filter" in place (intervening tissue and skin), allowing only the larger signals to be discernible. Also due to the muscle and tissue, the noise levels is higher in the surface ECG compared to the EGM signals.

As compared to the 12-lead ECG, which provides a three-dimensional view of the heart's polarization cycle, only a one or two-dimensional view is provided by the wireless ECG and a one dimensional view is provided by the EGM sensors. It is recognized by the medical community that the amount of information increases as the number of (independent) dimensions is increased from one to three. However, the 12-lead ECG is not portable and cannot be used for continuous monitoring without hindering the patient's daily activities.

The paper presents an innovative Cardiac Rhythm Management (*iCRM*) System and is organized as follows. In section II, we present a literature review of the technologies used for the *i*CRM. In section III we describe the *i*CRM. Section IV presents the simulation setup and preliminary results of our approach. Finally, we present our conclusions and future research directions in section V.

## II. LITERATURE REVIEW

### A. Implanted Devices

Implanted Cardiac Rhythm Management devices are used to monitor and correct arhythmia. There are three classes of Implantable Cardiac Rhythm Management devices: 1) Pacemakers; 2) Implantable Cardioverter/Defibrillators (ICD); and 3) Cardiac Resynchronization Therapy devices (CRT). The respective functions of these devices, which include sensing and actuating, vary depending on the type of arrythmia. These devices are implanted in the patient's chest area close to the collarbone. Special conductive wires [catheters] extending from the device are passed through the blood vessels and are placed into the heart. The sensors on the tip of the catheters measure the fluctuation of voltage in the locus of cells that it is in contact with and sends this information to the implanted device. The device, then administers voltage [actuation]. This informations is called an Intracardiac electrogram EGM [2]. The EGM is a fine-grained view of the heart's polarization cycle. The implanted device processes and analyzes this information to monitor the heart condition. Based on this information an appropriate voltage is applied depending on the arrhythmia. Pacemakers apply a small voltage to bring the heart rate back to normal. ICD/CRT devices apply a certain amount of shock (voltage) to the heart to bring it out of potentially fatal arrhythmia back to normal heart rhythm. Fig. 1 shows a basic block diagram of a ICD/CRM device [3].

### B. Learning Systems

Machine Learning [4] is a branch of artificial intelligence that enables a computer or an embedded system to detect data patterns from input data provided without being programmed, *a priori*, with explicit rules [as would be the case with an Expert System]. The leaning system develops rules based on labelled, data of a given process, provided in a training mode to classify data and effectively aids or replaces expert involvement.

Learning systems employ machine-learning algorithms on substantial records of training data of a given process to learn patterns and predict outcomes. The success of the learning system depends on many factors such as, the type of algorithm chosen, nature of the data, quality of the data, etc. Success is measured using different metrics such as confusion matrices, receiver operator characteristics (ROC), comparison of true and false positives, confidence measures, etc.

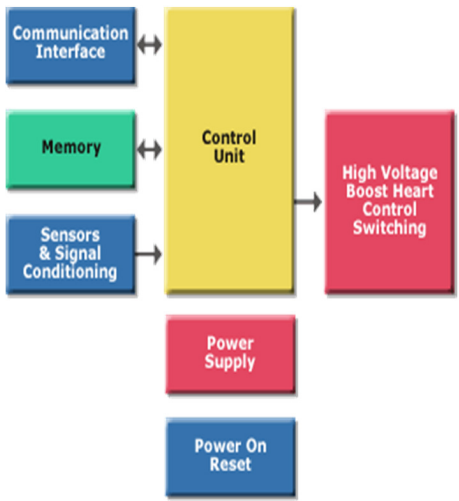

Fig. 1. Block Diagram of ICD/CRT Architecture.

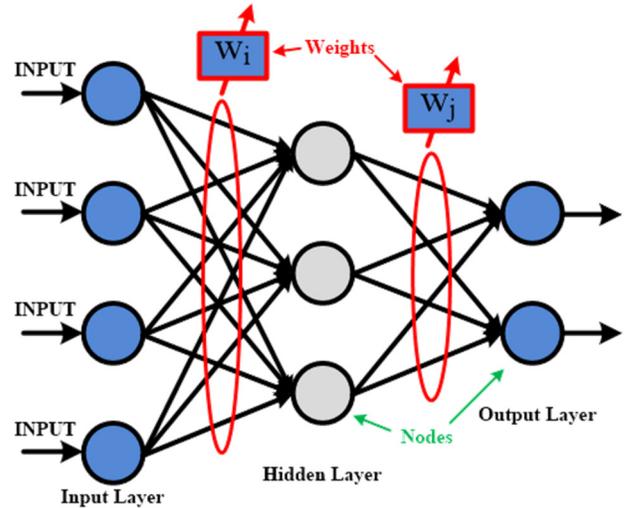

Fig. 2. Architechture of a Neural Network

Typically, the data are divided into three phases: (1) training data, (2) validation data and (3) testing data. The training data are used to teach the system and develop rules. Each outcome is labeled and the system knows this label. The validation data are used to perform preliminary testing on the algorithm. The testing data are used to perform a final test on the quality of the algorithm [5].

The ECG and EGM data is a set of voltage fluctuations in time that form patterns depending upon the condition of the heart. For example, in the case of atrial fibrillation the ECG and EGM could show a rate of over 300 peak fluctuations per minute (normal rhythm follows a pattern of 72 patterns per minute). The algorithm that we employ must be trained to recognize patterns and associate them with specific arrhythmias.

To achieve this goal we use a learning system technique called an Artificial Neural Network (ANN) in order to learn ECG and EGM patterns and correctly classify them. Artificial Neural Networks are based on the concept of biologically occurring neural networks that are present in the central nervous system and are responsible for human learning and computation. An ANN uses a plurality of nodes or 'neurons' [the circles in the figure] to recursively train and adapt weights [the *W*'s in the figure] to correctly recognize patterns and associated labels (classes) as shown in Fig. 2. ANN systems are especially useful in recognizing patterns and are used substantially in ECG analysis[3], [6]–[9].

## III. INNOVATIVE WIRELESS CARDIAC RHYTHM MANAGEMENT SYSTEM

The improved Cardiac Rhythm Management (*iCRM*) System, which is illustrated in Fig. 3, is designed to improve patient outcomes by using existing technologies (e.g., a wireless ambulatory ECG and an EGM) in combination with a novel

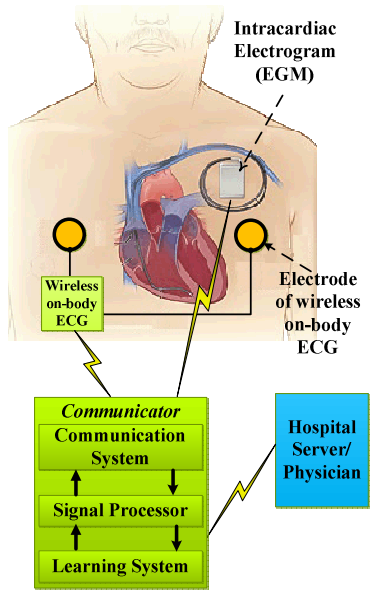

Fig. 3. Improved Cardiac Rhythm Management (*i*CRM) System Architecture.

*Communicator* to provide enhanced multi-dimensional networked ECG information. The networked information is input to a *Learning System* on-board the *Communicator* that detects the arrhythmia from the ECG/EGM patterns and provides enhanced actuating decisions to the implanted CRM devices [i.e., actuators] or relays accurate diagnoses to a physician, to create an improved CRM system.

The *Learning System* makes decisions based on the received information from several networked sensors (depending on the configuration of lead placements of the wireless ECG as well as the EGM) and supplies information to the implanted actuators and simultaneously sends information to the hospital/physician for round-the-clock and immediate/continuous monitoring. The *Communicator* features: (1) real-time processing, (2) a better grade of intelligence than available in the implants realized in the advanced learning techniques employed in the communicator (not present in present-day CRM), and (3) serves as the central decision-making and communication device to the sensor/actuator network.

In the case of ICDs, the proposed system adds value by aiding the EGM monitoring with the *Communicator* performing the critical decision-making role. From a fabrication point of view, this will also help to reduce complexity in the ICD because the decision making infrastructure is moved into the *Communicator*.

The preprocessing that the *Communicator* provides for the CRT may greatly reduce the complexity and size of the system, as it is one of the larger cardiac rhythm management (CRM) devices. The inclusion of the external ECG in the monitoring process helps to add an extra dimension in the monitoring process. This will help in acquiring better information for actuation. The implanted actuator (ICD or CRT) communicates with the *Communicator* and the *Communicator* receives the information from the 2- or 3-lead external ECG and the 2-lead ECG present inside the actuator. The information collected from these leads is fed into the *Learning System* to make more accurate decisions than is achievable by either device alone.

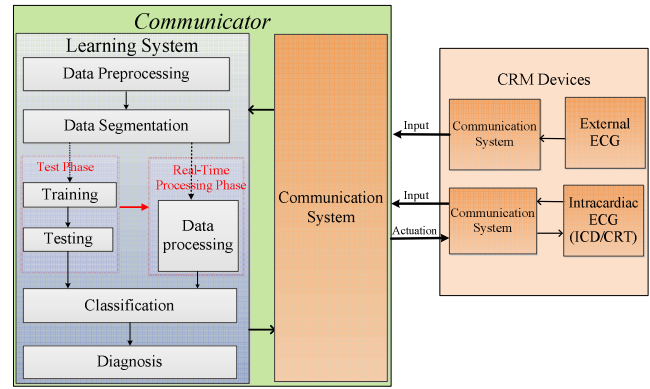

Fig. 4. Functional view of the *Communicator Learning Systems* algorithm.

Figure 4 shows the *Communicator* communicating with the monitor (ECGs) and actuator (ICD/CRT). The *Communicator* acts as a preprocessor for the actuator by processing the information from the monitors and making decisions that improve the accuracy and efficiency of the actuator. For example, the Holter monitor [10], which is a 2-lead on-body ECG, supplies information to the Communicator, and, in turn, it calls upon the expertise of the *Learning System* to make decisions for the benefit of the ICD/CRT (better pulsing accuracy).

To achieve low data rates for wireless communication, we include real-time compression inside the ECG monitors before transmitting the information to the *Communicator*. This increases the battery's lifetime of the monitor because fewer bits are transmitted. The information processed by the *Communicator* is also relayed on a cellular or Wi-Fi network, through the internet, to the hospital data center. Besides the data obtained in conventional ECG measurements, it also includes a reliable diagnosis and report of the recent ICD/CRT device activity. This leads to a more holistic understanding of the patient's problem and allows medical attention to be brought to the patient faster.

## IV. PRELIMINARY RESULTS

### A. Physiobank Database

In order to test the above mentioned hypothesis we used a database from Physiobank [11] that contained both surface ECG and EGM information that was simultaneously recorded. The EGM traces were obtained using a catheter containing 10 sensors. The catheter was positioned in different locations within the cardiac atrium, and information was recorded while the patient underwent an episode of either Atrial Fibrillation [AFB] or Atrial Flutter [AFL]. The data were digitized at 1 KHz and contained 8 signal traces for every patient, 5 EGM signals and 3 surface ECG signals recorded simultaneously. Another database containing surface ECGs of patients with a normal heart beat or Normal Sinus Rhythm [NSR], also from Physiobank [11], was adapted into our experiment.

## B. Results with Surface ECG

We present our preliminary results from passing the surface ECG signal through an ANN algorithm using the MATLAB neural network toolbox.

The surface ECG signals are divided into data segments of a constant time period. This time period was chosen to represent a standard PQRST signal of a normal heartbeat [10]. This amounted to about 106 samples after appropriate down sampling. These data segments were labeled according to the heart condition (Atrial Fibrillation, Atrial Flutter or Normal Sinus Rhythm) and input to the neural network using a standard back propagation algorithm [12]. The neural network had 3 layers, input layer, output layer and one hidden layer containing 10 neurons.

Figure 5 shows the *confusion matrices* [4], [5], [12] for training, testing and validation. The rows contain the true classes and the columns contain the class that was predicted by the algorithm. We can see that 99.2 percent of data segments were correctly classified. Fig. 6 shows the number of times the algorithm went through the entire data set and it's relationship with the mean squared error for training, validation and testing.

## C. Prototype System Design

The prototype system implementation of the *i*CRM involves that is shown in Fig. 6 consists of three components: the *Communicator* acting as bulk data processing and decision engine, and two device emulators, one to emulate the EGM and the other to emulate an external ECG Holter monitor.

The *Communicator* is implemented using a 32-bit Freescale K-70 microcontroller operating at 120MHz. To simplify hardware development and focus on proof of concept of the ANN algorithm in real-time, the Freescale Tower System TWRK70F120M is employed as a complete hardware development platform solution for experimentation with real-time behavior of the learning system algorithm.

Likewise, the device emulators are also built using the Freescale Tower System K-60 MCU operating at 100MHz on the TWRK60D100M development board. These devices contain a Bluetooth and Zigbee radio, to emulate the types of proprietary and non-proprietary radio interfaces found on commonly available implanted and external ECG devices. Each device is hard coded with data from the Physiobank database corresponding to EGM and ECG traces. The devices, when powered on, wait for a synchronization signal from the *Communicator* and then begin transmitting data using their respective radio. The synchronization is important to ensure the data being received by the *Communicator* is accurate between the two sources to the corresponding time value ECG sample.

A real-time operating system named MQX is provided by Freescale to support their MCU with handling task scheduling, peripheral management, and data flow within the device. MQX is a priority task-based system with an API programmable interface to communicate to external devices. The top-level tasks scheduled by MQX are to connect and maintain a wireless connection with the Bluetooth and Zigbee modems, read the input buffer to receive incoming ECG data, process the data using the *Learning System*, and transmit the decision result when available via the UART.

The core of the *Learning System* is an ANN subroutine based on C code that was tested on a PC and the results compared against Matlab simulation with the same training data and input data for verification of operation. For the purpose of the initial hardware emulation, the training data is pre-programmed as a fixed variable array that is configured during compile time. In future iterations, the *Learning System* and *Communicator* will allow dynamic training of the ANN to add additional decision results once data is available for conditions beyond the scope of the Physionet database. Also, the ECG device emulators will be configured to read data from a removable storage device to provide additional system testing using patient ECG data from other sources.

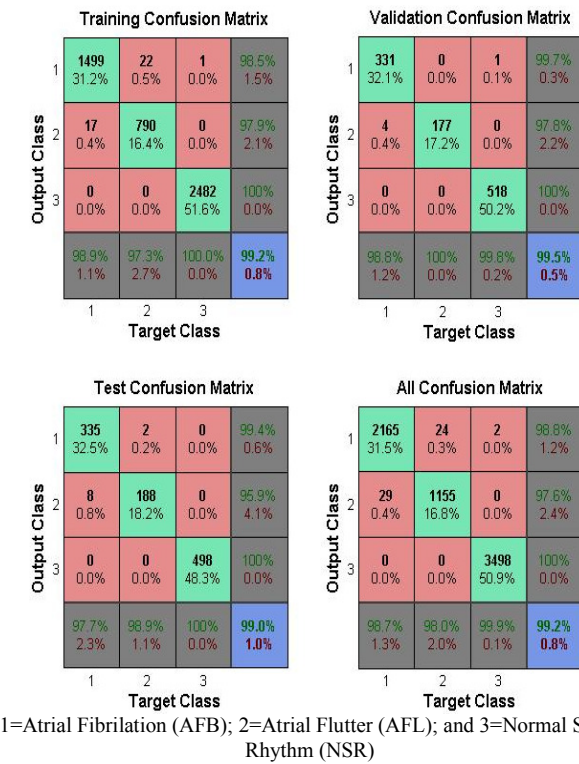

1=Atrial Fibrillation (AFB); 2=Atrial Flutter (AFL); and 3=Normal Sinus Rhythm (NSR)

Fig. 5. Confusion Matrix generated by the MATLAB Toolbox.

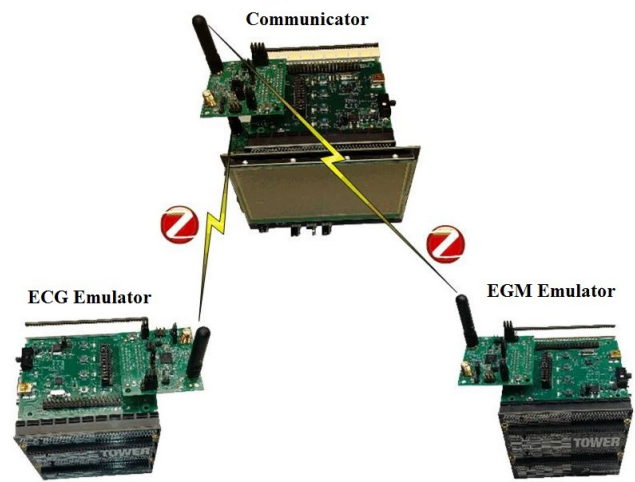

Fig. 6. *i*CRM prototype system with FreeScale equipment

The flexibility of the Tower System peripheral interface will only require minor adjustments in the *Communicator* hardware and software to incorporate any proprietary wireless technologies that may be used in commercially available ECG systems.

## V. CONCLUSIONS AND FUTURE DIRECTIONS

A high degree of accuracy was achieved in predicting atrial arrhythmia in 8 patients using only an external ECG when we used an ANN neural network with 10 hidden neurons with a back propagation algorithm.

Future work involves extracting information from the EGM that compliments the information of the surface ECG and implementing these algorithms on hardware. In addition we plan to explore the benefits of using the VCG representation by converting both the ECG and the EGM signals into the VCG format, $VCG_{EGM}$ and $VCG_{ECG}$ respectively, to see if this improves decision making of the learning system [13].

### ACKNOWLEDGMENT

This research was supported in part by Jabil Inc., NSF Grant IIP-1217306, the Florida 21st Century Scholars program and the Florida High Tech Corridor Matching Grants Research Program.